# Dimension in Complexity Classes

Jack H. Lutz*
Department of Computer Science
Iowa State University
lutz@cs.iastate.edu

**Abstract**

A theory of resource-bounded dimension is developed using *gales*, which are natural generalizations of martingales. When the resource bound $\Delta$ (a parameter of the theory) is unrestricted, the resulting dimension is precisely the classical Hausdorff dimension (sometimes called "fractal dimension"). Other choices of the parameter $\Delta$ yield internal dimension theories in E, $E_2$, ESPACE, and other complexity classes, and in the class of all decidable problems. In general, if $\mathcal{C}$ is such a class, then *every* set $X$ of languages has a dimension in $\mathcal{C}$, which is a real number $\dim(X \mid \mathcal{C}) \in [0,1]$. Along with the elements of this theory, two preliminary applications are presented:

1. For every real number $0 \leq \alpha \leq \frac{1}{2}$, the set $\mathrm{FREQ}(\leq \alpha)$, consisting of all languages that asymptotically contain at most $\alpha$ of all strings, has dimension $\mathcal{H}(\alpha)$ — the binary entropy of $\alpha$ — in E and in $E_2$.

2. For every real number $0 \leq \alpha \leq 1$, the set $\mathrm{SIZE}(\alpha \frac{2^n}{n})$, consisting of all languages decidable by Boolean circuits of at most $\alpha \frac{2^n}{n}$ gates, has dimension $\alpha$ in ESPACE.

## 1 Introduction

Since the development of resource-bounded measure in 1991 [9], the investigation of the internal, measure-theoretic structure of complexity classes has produced a rapidly growing body of new insights and results. As indicated by the survey papers [1, 3, 10, 12], this line of inquiry has shed light on a wide variety of topics in computational complexity. The ongoing fruitfulness of this research is not surprising because resource-bounded measure is a complexity-theoretic generalization of classical Lebesgue measure, which was one of the most powerful quantitative tools of twentieth-century mathematics.

In spite of this power, there are certain inherent limitations to the amount of quantitative information that resource-bounded measure can provide in computational complexity. One of these limitation arises from the resource-bounded Kolmogorov zero-one law, which was proven by Lutz [11] and has recently been strengthened by Dai [4]. For any class $\mathcal{C}$ in which resource-bounded measure is defined, and for any set $X$ of languages that is — like most sets of interest in computational complexity — closed under finite variations, the zero-one law says that the measure of $X$ in $\mathcal{C}$ must be 0 or 1 or undefined. A second limitation arises from the simple fact that even a measure 0 subset of a complexity class may have internal structure that we would like to elucidate quantitatively. Of course both these limitations were already present in classical Lesbesgue measure theory.

---

*This research was supported in part by National Science Foundation Grants 9610461 and 9988483.

In 1919, Hausdorff [7] augmented classical Lesbesgue measure theory with a theory of dimension. This theory assigns to *every* subset $X$ of a given metric space a real number $\dim_H(X)$, which is now called the *Hausdorff dimension* of $X$. In this paper, we are interested in the case where the metric space is the Cantor space $\mathbf{C}$, consisting of all decision problems (i.e., all languages $A \subseteq \{0,1\}^*$). In this case the Hausdorff dimension of a set $X \subseteq \mathbf{C}$ (which is defined precisely in section 3) is a real number $\dim_H(X) \in [0,1]$. The Hausdorff dimension is monotone, with $\dim_H(\emptyset) = 0$ and $\dim_H(\mathbf{C}) = 1$. Moreover, if $\dim_H(X) < \dim_H(\mathbf{C})$, then $X$ is a measure 0 subset of $\mathbf{C}$. Hausdorff dimension thus overcomes both of the limitations mentioned in the preceding paragraph.

In this paper we develop *resource-bounded dimension*, which is a complexity-theoretic generalization of classical Hausdorff dimension. We carry out this generalization in two steps. We first prove a new characterization of classical Hausdorff dimension in terms of *gales*, which are a natural generalization of the martingales that are the basis of resource-bounded measure. (Our characterization can be regarded as an analog of Ville's martingale characterization of the Lesbesgue measure 0 sets [19].) We then generalize classical dimension by introducing a resource bound $\Delta$ (a parameter of the theory) and requiring the gales to be $\Delta$-computable. We show that this induces a well-behaved notion of dimension in the corresponding class $R(\Delta)$, which is defined exactly as in resource-bounded measure. We write $\dim(X \mid R(\Delta))$ for the dimension of the set $X$ in the class $R(\Delta)$. If $\Delta$ is unrestricted, then $R(\Delta) = \mathbf{C}$ and $\dim(X \mid R(\Delta)) = \dim(X \mid \mathbf{C})$ is precisely $\dim_H(X)$. However, other choices of $\Delta$ allow $R(\Delta)$ to be interesting complexity classes, in which case $\dim(X \mid R(\Delta))$ is a quantitative measure of the dimension of $X \cap R(\Delta)$ *as a subset of $R(\Delta)$*.

After presenting the elements of resource-bounded dimension, we present two preliminary applications of the theory. First, for each real number $\alpha \in [0,1]$, let $\text{FREQ}(\leq \alpha)$ be the set of all languages that asymptotically contain at most $\alpha$ of all strings. (This set is defined precisely in section 5.) We prove that, for every real number $\alpha \in [0, \frac{1}{2}]$,

$$\dim(\text{FREQ}(\leq \alpha) \mid \text{E}) = \mathcal{H}(\alpha)$$

and

$$\dim(\text{FREQ}(\leq \alpha) \mid \text{E}_2) = \mathcal{H}(\alpha) \ ,$$

where $\text{E} = \text{DTIME}(2^{\text{linear}})$, $\text{E}_2 = \text{DTIME}(2^{\text{poly}})$, and $\mathcal{H}$ is the binary entropy function of Shannon information theory.

Our second application concerns Boolean circuit-size complexity in the complexity class $\text{ESPACE} = \text{DSPACE}(2^{\text{linear}})$. For each real number $\alpha \in [0,1]$, let $\text{SIZE}(\alpha \frac{2^n}{n})$ be the set of all languages that can be decided by Boolean circuits consisting of at most $\alpha \frac{2^n}{n}$ gates. We prove that for all $\alpha \in [0,1]$,

$$\dim(\text{SIZE}(\alpha \frac{2^n}{n}) \mid \text{ESPACE}) = \alpha \ .$$

These applications are interesting because they show that resource-bounded dimension interacts informatively with information theory and Boolean circuit-size complexity. However, they are clearly only the beginning. Classical Hausdorff dimension is a sophisticated mathematical theory that has emerged as one of the most important tools for the investigation of fractal sets. (See, for example [6] for a good introduction and overview.) Many sets of interest in computational complexity seem to have "fractal-like" structures. Resource-bounded dimension will be a useful tool for the study of such sets.

## 2 Preliminaries

A *decision problem* (a.k.a. *language*) is a set $A \subseteq \{0,1\}^*$. We identify each language with its characteristic sequence $[\![s_0 \in A]\!][\![s_1 \in A]\!][\![s_2 \in A]\!] \cdots$, where $s_0, s_1, s_2, \ldots$ is the standard enumer-ation of $\{0,1\}^*$ and $[\![\phi]\!] = \texttt{if } \phi \texttt{ then } 1 \texttt{ else } 0$. We write $A[i..j]$ for the string consisting of the $i$-th through $j$-th bits of (the characteristic sequence of) $A$. The Cantor space $\mathbf{C}$ is the set of all decision problems.

If $w \in \{0,1\}^*$ and $x \in \{0,1\}^* \cup \mathbf{C}$, then $w \sqsubseteq x$ means that $w$ is a prefix of $x$, and $w \sqsubsetneq x$ means that $w$ is a proper prefix of $x$. The *cylinder* generated by a string $w \in \{0,1\}^*$ is $\mathbf{C}_w = \{A \in \mathbf{C} \mid w \sqsubseteq A\}$.

A *prefix set* is a language $A$ such that no element of $A$ is a prefix of any other element of $A$.

If $A$ is a language and $n \in \mathbb{N}$, then we write $A_{=n} = A \cap \{0,1\}^n$ and $A_{\leq n} = A \cap \{0,1\}^{\leq n}$.

All logarithms in this paper are base 2.

For each $i \in \mathbb{N}$ we define a class $G_i$ of functions from $\mathbb{N}$ into $\mathbb{N}$ as follows.

$$\begin{aligned} G_0 &= \{f \mid (\exists k)(\forall^\infty n) f(n) \leq kn\} \\ G_{i+1} &= 2^{G_i(\log n)} = \{f \mid (\exists g \in G_i)(\forall^\infty n) f(n) \leq 2^{g(\log n)}\} \end{aligned}$$

We also define the functions $\hat{g}_i \in G_i$ by $\hat{g}_0(n) = 2n$, $\hat{g}_{i+1}(n) = 2^{\hat{g}_i(\log n)}$. We regard the functions in these classes as growth rates. In particular, $G_0$ contains the linearly bounded growth rates and $G_1$ contains the polynomially bounded growth rates. It is easy to show that each $G_i$ is closed under composition, that each $f \in G_i$ is $o(\hat{g}_{i+1})$, and that each $\hat{g}_i$ is $o(2^n)$. Thus $G_i$ contains superpolynomial growth rates for all $i > 1$, but all growth rates in the $G_i$-hierarchy are subexponential.

Within the class DEC of all decidable languages, we are interested in the exponential complexity classes $\mathrm{E}_i = \mathrm{DTIME}(2^{G_{i-1}})$ and $\mathrm{E}_i\mathrm{SPACE} = \mathrm{DSPACE}(2^{G_{i-1}})$ for $i \geq 1$. The much-studied classes $\mathrm{E} = \mathrm{E}_1 = \mathrm{DTIME}(2^{\text{linear}})$, $\mathrm{E}_2 = \mathrm{DTIME}(2^{\text{polynomial}})$, and $\mathrm{ESPACE} = \mathrm{E}_1\mathrm{SPACE} = \mathrm{DSPACE}(2^{\text{linear}})$ are of particular interest.

We use the following classes of functions.

$$\begin{aligned} \text{all} &= \{f \mid f : \{0,1\}^* \to \{0,1\}^*\} \\ \text{comp} &= \{f \in \text{all} \mid f \text{ is computable}\} \\ \mathrm{p}_i &= \{f \in \text{all} \mid f \text{ is computable in } G_i \text{ time}\} \ (i \geq 1) \\ \mathrm{p}_i\text{space} &= \{f \in \text{all} \mid f \text{ is computable in } G_i \text{ space}\} \ (i \geq 1) \end{aligned}$$

(The length of the output *is* included as part of the space used in computing $f$.) We write p for $\mathrm{p}_1$ and pspace for $\mathrm{p}_1$space. Throughout this paper, $\Delta$ and $\Delta'$ denote one of the classes all, comp, $\mathrm{p}_i(i \geq 1)$, $\mathrm{p}_i\text{space}(i \geq 1)$.

A *constructor* is a function $\delta : \{0,1\}^* \to \{0,1\}^*$ that satisfies $x \sqsubsetneq \delta(x)$ for all $x$. The *result* of a constructor $\delta$ (i.e., the language *constructed* by $\delta$) is the unique language $R(\delta)$ such that $\delta^n(\lambda) \sqsubseteq R(\delta)$ for all $n \in \mathbb{N}$. Intuitively, $\delta$ constructs $R(\delta)$ by starting with $\lambda$ and then iteratively generating successively longer prefixes of $R(\delta)$. We write $R(\Delta)$ for the set of languages $R(\delta)$ such that $\delta$ is a constructor in $\Delta$. The following facts are the reason for our interest in the above-defined classes of functions.

$R(\text{all}) = \mathbf{C}$.
$R(\text{comp}) = \mathrm{DEC}$.
For $i \geq 1$, $R(\mathrm{p}_i) = \mathrm{E}_i$.
For $i \geq 1$, $R(\mathrm{p}_i\text{space}) = \mathrm{E}_i\mathrm{SPACE}$.

If $D$ is a discrete domain, then a function $f : D \longrightarrow [0, \infty)$ is $\Delta$-*computable* if there is a function $\hat{f} : \mathbb{N} \times D \longrightarrow \mathbb{Q} \cap [0, \infty)$ such that $|\hat{f}(r, x) - f(x)| \leq 2^{-r}$ for all $r \in \mathbb{N}$ and $x \in D$ and $\hat{f} \in \Delta$ (with $r$ coded in unary and the output coded in binary). We say that $f$ is *exactly* $\Delta$-*computable* if $f : D \longrightarrow \mathbb{Q} \cap [0, \infty)$ and $f \in \Delta$.

## 3 Gales and Hausdorff Dimension

In this section we introduce gales and supergales, which are a generalization of martingales and supermartingales, and use these to give a new characterization of classical Hausdorff dimension.

**Definition.** *Let $s \in [0, \infty)$.*

1. *An $s$-supergale is a function $d : \{0,1\}^* \longrightarrow [0, \infty)$ that satisfies the condition*

$$d(w) \geq 2^{-s}[d(w0) + d(w1)] \quad (3.1)$$

   *for all $w \in \{0,1\}^*$*

2. *An $s$-gale is an $s$-supergale that satisfies (3.1) with equality for all $w \in \{0,1\}^*$.*

3. *A* supermartingale *is a 1-supergale.*

4. *A* martingale *is a 1-gale.*

**Lemma 3.1.** *Let $s \in [0, \infty)$. If $d$ is an $s$-supergale and $B \subseteq \{0,1\}^*$ is a prefix set, then for all $w \in \{0,1\}^*$,*

$$\sum_{u \in B} 2^{-s|u|} d(wu) \leq d(w) \ .$$

***Proof:*** We first use induction on $n$ to prove that for all $n \in \mathbb{N}$, the lemma holds for all prefix sets $B \subseteq \{0,1\}^{\leq n}$. For $n = 0$, this is trivial. Assume that it holds for $n$, and let $A \subseteq \{0,1\}^{\leq n+1}$ be a prefix set. Let

$$A' = \{u \in \{0,1\}^n \mid u0 \in A \text{ or } u1 \in A\} \ ,$$

and let

$$B = A_{\leq n} \cup A' \ .$$

Note that $B$ is a prefix set and $A_{\leq n} \cap A' = \emptyset$ (because $A$ is a prefix set). Also, for all $w \in \{0,1\}^*$,

$$\sum_{u \in A_{=n+1}} 2^{-s|u|} d(wu)$$
$$= 2^{-s(n+1)} \sum_{u \in A_{=n+1}} d(wu)$$
$$\leq 2^{-s(n+1)} \sum_{u \in A'} [d(wu0) + d(wu1)]$$
$$\leq 2^{-s(n+1)} \sum_{u \in A'} 2^s d(wu)$$
$$= \sum_{u \in A'} 2^{-s|u|} d(wu) \ .$$

Since $B \subseteq \{0,1\}^{\leq n}$, it follows by the induction hypothesis that for all $w \in \{0,1\}^*$,

$$\sum_{u \in A} 2^{-s|u|} d(wu)$$
$$= \sum_{u \in A_{\leq n}} 2^{-s|u|} d(wu) + \sum_{u \in A_{=n+1}} 2^{-s|u|} d(wu)$$
$$\leq \sum_{u \in A_{\leq n}} 2^{-s|u|} d(wu) + \sum_{u \in A'} 2^{-s|u|} d(wu)$$
$$= \sum_{u \in B} 2^{-s|u|} d(wu)$$
$$\leq d(w) \ .$$

This completes the proof that for all $n \in \mathbb{N}$, the lemma holds for all prefix sets $B \subseteq \{0,1\}^{\leq n}$.

To complete the proof of the lemma, let $B$ be an arbitrary prefix set. Then for all $w \in \{0,1\}^*$,

$$\sum_{u \in B} 2^{-s|u|} d(wu) = \sup_{n \in \mathbb{N}} \sum_{u \in B_{\leq n}} 2^{-s|u|} d(wu) \leq d(w) \ .$$

□

**Corollary 3.2.** *Let $s \in [0,\infty)$, $0 < \alpha \in \mathbb{R}$, and $w \in \{0,1\}^*$. If $d$ is an $s$-supergale such that $d(w) > 0$ and $B \subseteq \{0,1\}^*$ is a prefix set such that $d(wu) \geq \alpha 2^{(s-1)|u|} d(w)$ for all $u \in B$, then*

$$\sum_{u \in B} 2^{-|u|} \leq \frac{1}{\alpha} \ .$$

*Proof:* Assume the hypothesis. Then by Lemma 3.1,

$$d(w) \geq \sum_{u \in B} 2^{-s|u|} d(wu) \geq \alpha d(w) \sum_{u \in B} 2^{-|u|} \ ,$$

whence the corollary follows. □

**Corollary 3.3.** *Let $d$ be an $s$-supergale, where $s \in [0,1]$. Then for all $w \in \{0,1\}^*$, $l \in \mathbb{N}$, and $0 < \alpha \in \mathbb{R}$, there are fewer than $\frac{2^l}{\alpha}$ strings $u \in \{0,1\}^l$ for which*

$$\max_{v \sqsubseteq u} 2^{(1-s)|v|} d(wv) > \alpha d(w).$$

*In particular, there is at least one string $u \in \{0,1\}^l$ such that $d(wv) \leq 2^{(s-1)|v|} d(w)$ for all $v \sqsubseteq u$.*

*Proof:* Let $d$, $s$, $w$, $l$ and $\alpha$ be as given, and let

$$A = \{u \in \{0,1\}^l \mid \max_{v \sqsubseteq u} 2^{(1-s)|v|} d(wv) > \alpha d(w)\}.$$

If $A = \emptyset$ the corollary is trivially affirmed, so assume that $A \neq \emptyset$. (Note that this implies $d(w) > 0$.) Let $B$ be the set of all $v \in \{0,1\}^{\leq l}$ such that $2^{(1-s)|v|} d(wv) > \alpha d(w)$ but $2^{(1-s)|v'|} d(wv') \leq \alpha d(w)$ for all $v' \sqsubsetneq v$. Then $B$ is a prefix set, and

$$A = \{u \in \{0,1\}^l \mid (\exists v \sqsubseteq u) v \in B\},$$

so
$$|A| = \sum_{v \in B} 2^{l-|v|} = 2^l \sum_{v \in B} 2^{-|v|}.$$

Let
$$\alpha' = \min_{v \in B} 2^{(1-s)|v|} \frac{d(wv)}{d(w)},$$

and note that $\alpha < \alpha' < \infty$. Then $B$ is a prefix set such that $d(wv) \geq \alpha' 2^{(s-1)|v|} d(w)$ for all $v \in B$, so Corollary 3.2 tells us that
$$|A| = 2^l \sum_{v \in B} 2^{-|v|} \leq \frac{2^l}{\alpha'} < \frac{2^l}{\alpha}.$$

This proves the main assertion of the corollary. The last sentence of the corollary follows immediately by taking $\alpha = 1$. □

**Corollary 3.4.** *If $d$ is an $s$-supergale, where $s \in [0, \infty)$, then for all $w, u \in \{0,1\}^*$,*
$$d(wu) \leq 2^{s|u|} d(w) .$$

**Proof:** Let $d$, $s$, $w$, and $u$ be as given, and let $l = |u|$. Let $\beta > 2^{s|u|}$ be arbitrary. It suffices to show that $d(wu) \leq \beta d(w)$.

Let $\alpha = \beta 2^{(1-s)l}$. Then, for all $v \in \{0,1\}^l$,
$$d(wv) > \beta d(w) \Leftrightarrow d(wv) > \alpha 2^{(s-1)l} d(w) ,$$

so Corollary 3.3 tells us that there are fewer than $\frac{2^l}{\alpha}$ strings $v \in \{0,1\}^l$ for which $d(wv) > \beta d(w)$. Since $\frac{2^l}{\alpha} = \frac{2^{s|u|}}{\beta} < 1$, it follows that $d(wu) \leq \beta d(w)$. □

**Observation 3.5.** *Let $s \in [0, \infty)$. For each $k \in \mathbb{N}$, let $d_k$ be an $s$-gale, and let $a_k \in [0, \infty)$.*

1. *For each $n \in \mathbb{Z}^+$, $\sum_{k=0}^{n-1} a_k d_k$ is an $s$-gale.*

2. *If $\sum_{k=0}^{\infty} a_k d_k(\lambda) < \infty$, then $\sum_{k=0}^{\infty} a_k d_k$ is an $s$-gale.*

**Definition.** *Let $d$ be an $s$-supergale, where $s \in [0, \infty)$.*

1. *We say that $d$* succeeds *on a language $A \in \mathbf{C}$ if*
$$\limsup_{n \to \infty} d(A[0 \ldots n-1]) = \infty .$$

2. *The* success set *of $d$ is*
$$S^\infty[d] = \{A \in \mathbf{C} \mid d \text{ succeeds on } A\} .$$

We now review the classical definition of Hausdorff dimension. Since we are primarily interested in the computational complexities of decision problems, we focus on Hausdorff dimension in the Cantor space $\mathbf{C}$.

For each $k \in \mathbb{N}$, we let $\mathcal{A}_k$ be the collection of all prefix sets $A$ such that $A_{<k} = \emptyset$. For each $X \subseteq \mathbf{C}$, we then let
$$\mathcal{A}_k(X) = \{A \in \mathcal{A}_k \mid X \subseteq \bigcup_{w \in A} \mathbf{C}_w\} .$$

If $A \in \mathcal{A}_k(X)$, then we say that the prefix set $A$ *covers* the set $X$. For $X \in \mathbf{C}$, $s \in [0, \infty)$, and $k \in \mathbb{N}$, we then define
$$H_k^s(X) = \inf_{A \in \mathcal{A}_k(X)} \sum_{w \in A} 2^{-s|w|} .$$

**Digression.** Readers familiar with Hausdorff dimension may prefer to regard it as arising from a measure or, more commonly, a metric on the underlying space. If we regard Hausdorff dimension as arising from a measure on $\mathbf{C}$, then the term $2^{-s|w|}$ in the above sum in interpreted as $\mu(\mathbf{C}_w)^s$, where $\mu(\mathbf{C}_w) = 2^{-|w|}$ is the Lebesgue measure of the cylinder $\mathbf{C}_w$. If we instead regard Hausdorff dimension as arising from a metric on $\mathbf{C}$, then the metric is

$$d(A, B) = 2^{-\min\{n \in \mathbb{N} | A[n] \neq B[n]\}}$$

(where $\min \emptyset = \infty$ so $d(A, A) = 0$), and the term $2^{-s|w|}$ is interpreted as $\mathrm{diam}(\mathbf{C}_w)^s$, where

$$\mathrm{diam}(X) = \sup_{A, B \in X} d(A, B)$$

is the *diameter* of a set $X \subseteq \mathbf{C}$. Such interpretations may provide context, but our technical development does not use them.

**Definition.** *For $s \in [0, \infty)$ and $X \subseteq \mathbf{C}$, the s-dimensional Hausdorff outer measure of $X$ is*

$$H^s(X) = \lim_{k \to \infty} H^s_k(X) \ .$$

Since $H^s_k(X)$ is nondecreasing in $k$, this limit $H^s(X)$ exists, though it may be infinite. In fact, it is well-known that for every set $X \subseteq \mathbf{C}$, there is a real number $s^* \in [0, 1]$ with the following two properties.

(i) For all $0 \leq s < s^*$, $H^s(X) = \infty$.

(ii) For all $s > s^*$, $H^s(X) = 0$.

As the following definition states, this number $s^*$ is the Hausdorff dimension of $X$.

**Definition.** *The* Hausdorff dimension *of a set $X \subseteq \mathbf{C}$ is*

$$\dim_H(X) = \inf\{s \in [0, \infty) \mid H^s(X) = 0\} \ .$$

**Notation.** For $X \subseteq \mathbf{C}$, let $\mathcal{G}(X)$ be the set of all $s \in [0, \infty)$ such that there is an $s$-gale $d$ for which $X \subseteq S^\infty[d]$.

The following theorem gives a new characterization of classical Hausdorff dimension.

**Theorem 3.6.** (Gale Characterization of Hausdorff Dimension). *For all $X \subseteq \mathbf{C}$, $\dim_H(X) = \inf \mathcal{G}(X)$.*

***Proof:*** It suffices to show that for all $s \in [0, \infty)$,

$$H^s(X) = 0 \Leftrightarrow s \in \mathcal{G}(X) \ .$$

First, assume that $H^s(X) = 0$. Then $H^s_0(X) = 0$, which implies that for each $r \in \mathbb{N}$, there is a prefix set $A_r \in \mathcal{A}_0(X)$ such that $\sum_{w \in A_r} 2^{-s|w|} \leq 2^{-r}$. For each $r \in \mathbb{N}$, then, fix such a prefix set $A_r$, and define a function $d_r : \{0, 1\}^* \longrightarrow [0, \infty)$ as follows. Let $w \in \{0, 1\}^*$. If there exists $v \sqsubseteq w$ such that $v \in A_r$, then $d_r(w) = 2^{(s-1)(|w|-|v|)}$. Otherwise,

$$d_r(w) = \sum_{\substack{u \\ wu \in A_r}} 2^{-s|u|} \ .$$

It is routine to verify that the following conditions hold for all $r \in \mathbb{N}$.

(i) $d_r$ is an $s$-gale.

(ii) $d_r(\lambda) \leq 2^{-r}$.

(iii) For all $w \in A_r$, $d_r(w) = 1$.

Let $d = \sum_{r=0}^{\infty} 2^r d_{2r}$. By Observation 3.5, $d$ is an $s$-gale. To see that $X \subseteq S^\infty[d]$, let $B \in X$, and let $r \in \mathbb{N}$ be arbitrary. Since the prefix set $A_{2r}$ covers $X$, there exists $w \in A_{2r}$ such that $w \sqsubseteq B$. Then by (iii) above, $d(w) \geq 2^r d_{2r}(w) = 2^r$. Since $r \in \mathbb{N}$ is arbitrary, this shows that $B \in S^\infty$, confirming that $X \subseteq S^\infty[d]$

We have now shown that $d$ is an $s$-gale such that $X \subseteq S^\infty[d]$, whence $s \in \mathcal{G}(X)$.

Conversely, assume that $s \in \mathcal{G}(X)$. To see that $H^s(X) = 0$, let $k, r, \in \mathbb{N}$. It suffices to show that $H^s(X) \leq 2^{-r}$. If $X = \emptyset$ this is trivial, so assume that $X \neq \emptyset$.

Since $s \in \mathcal{G}(X)$, there is an $s$-gale $d$ such that $X \subseteq S^\infty[d]$. Note that $d(\lambda) > 0$ because $X \neq 0$. Let
$$a = 1 + \max\{d(w) \mid w \in \{0,1\}^{\leq k}\},$$
and let
$$A = \{w \in \{0,1\}^* \mid d(w) \geq 2^r a \text{ and } (\forall v) \, [v \sqsubsetneq w \Rightarrow d(v) < 2^r a]\}.$$

It is clear that $A$ is a prefix set with $A_{<k} = \emptyset$, so $A \in \mathcal{A}_k$. It is also clear that
$$X \subseteq S^\infty[d] \subseteq \bigcup_{w \in A} \mathbf{C}_w,$$
whence $A \in \mathcal{A}_k(X)$. By Lemma 3.1 and the definition of $A$, we have
$$d(\lambda) \geq \sum_{w \in A} 2^{-s|w|} d(w) \geq 2^r d(\lambda) \sum_{w \in A} 2^{-s|w|}.$$

Since $A \in \mathcal{A}_k(X)$ and $d(\lambda) > 0$, it follows that
$$H^s_k(X) \leq \sum_{w \in A} 2^{-s|w|} \leq 2^{-r}.$$

□

It is clear from the proof of Theorem 3.6 that the $s$-dimensional Hausdorff outer measure $H^s(X)$ can also be characterized in terms of $s$-gales, but we refrain from elaborating here.

## 4 Dimension in Complexity Classes

Motivated by the gale characterization of classical Hausdorff dimension, we now use resource-bounded gales to develop dimension in complexity classes. As with resource-bounded measure [9], our development contains a parameter $\Delta$ — the resource bound — which may be any one of the classes all, comp, p, pspace, $p_2$, $p_2$space, etc., defined in section 2.

**Notation.** For $X \subseteq \mathbf{C}$, let $\mathcal{G}_\Delta(X)$ be the set of all $s \in [0, \infty)$ such that there is a $\Delta$-computable $s$-gale $d$ for which $X \subseteq S^\infty[d]$.

**Definition.** *Let $X \subseteq \mathbf{C}$.*

1. *The $\Delta$-dimension of $X$ is $\dim_\Delta(X) = \inf \mathcal{G}_\Delta(X)$.*

2. *The dimension of $X$ in $R(\Delta)$ is $\dim(X \mid R(\Delta)) = \dim_\Delta(X \cap R(\Delta))$.*

Note that $\dim_\Delta(X)$ and $\dim(X \mid R(\Delta))$ are defined for *every* set $X \subseteq \mathbf{C}$. We call $\dim_{\text{comp}}(X)$ and $\dim_{\text{p}}(X)$ the *computable dimension* of $X$ and the *feasible dimension* of $X$, respectively.

The following observations are elementary but useful.

**Observations 4.1.**

1. *For all $X \subseteq Y \subseteq \mathbf{C}$,*
$$\dim_\Delta(X) \leq \dim_\Delta(Y)$$
*and*
$$\dim(X \mid R(\Delta)) \leq \dim(Y \mid R(\Delta)) .$$

2. *If $\Delta$ and $\Delta'$ are resource bounds such that $\Delta \subseteq \Delta'$, then for all $X \subseteq \mathbf{C}$,*
$$\dim_{\Delta'}(X) \leq \dim_\Delta(X) .$$

3. *For all $X \subseteq \mathbf{C}$, $0 \leq \dim(X \mid R(\Delta)) \leq \dim_\Delta(X) \leq 1$.*

4. *For all $X \subseteq \mathbf{C}$, $\dim(X \mid \mathbf{C}) = \dim_{\text{all}}(X) = \dim_H(X)$.*

5. *For all $X \subseteq \mathbf{C}$,*
$$\dim_\Delta(X) < 1 \Rightarrow \mu_\Delta(X) = 0$$
*and*
$$\dim(X \mid R(\Delta)) < 1 \Rightarrow \mu(X \mid R(\Delta)) = 0 .$$

*Proof:* Observations 1, 2, 4, and 5 follow immediately from the definitions. For observation 3, it suffices to show that $\dim_\Delta(\mathbf{C}) \leq 1$. For this, fix $s > 1$ such that $2^s$ is rational. Then the function

$$d : \{0,1\}^* \to [0, \infty)$$
$$d(w) = 2^{(s-1)|w|}$$

is a $\Delta$-computable $s$-gale such that $S^\infty[d] = \mathbf{C}$, so $s \in \mathcal{G}_\Delta(\mathbf{C})$, so $\dim_\Delta(\mathbf{C}) \leq s$. Since this holds for all $s > 1$ with $2^s$ rational, it follows that $\dim_\Delta(\mathbf{C}) \leq 1$. □

The fifth observation above shows that resource-bounded dimension offers a quantitative classification of sets that have resource-bounded measure 0. However, it should be emphasized that this classification is in terms of dimension, which is very different from measure. For example, Lemma 4.8 and its corollaries exhibit properties of dimension that contrast sharply with those of measure.

To proceed further, we need a little bit of technical machinery concerning the computability of gales and supergales.

**Observation 4.2.** *If $d$ is a $\Delta$-computable $s$-gale, then the real number $2^s$ is $\Delta$-computable.*

**Lemma 4.3.** *If $d$ is a $\Delta$-computable $s$-supergale and $2^s$ is $\Delta$-computable, then there is a $\Delta$-computable $s$-gale $\tilde{d}$ such that $S^\infty[d] \subseteq S^\infty[\tilde{d}]$. Moreover, if $d$ is exactly $\Delta$-computable and $2^s$ is rational, then $\tilde{d}$ is exactly $\Delta$-computable.*

**Proof:** Assume the hypothesis. Define

$$\tilde{d} : \{0,1\}^* \longrightarrow [0,\infty)$$

$$\tilde{d}(\lambda) = d(\lambda)$$

$$\tilde{d}(w0) = \frac{1}{2}[2^s \tilde{d}(w) + d(w0) - d(w1)]$$

$$\tilde{d}(w1) = \frac{1}{2}[2^s \tilde{d}(w) - d(w0) + d(w1)].$$

Then $\tilde{d}$ is clearly a $\Delta$-computable $s$-gale, and an easy induction shows that $\tilde{d}(w) \geq d(w)$ for all $w \in \{0,1\}^*$, whence $S^\infty[d] \subseteq S^\infty[\tilde{d}]$. Moreover, if $d$ is exactly $\Delta$-computable and $2^s$ is rational, then it is clear that $\tilde{d}$ is exactly $\Delta$-computable. $\square$

**Observation 4.4.** *If $d$ is an $s$-supergale and $s' \geq s$, then $d$ is an $s'$-supergale.*

**Corollary 4.5.** *Let $X \subseteq \mathbf{C}$ and $s' \geq s \geq 0$. If $s \in \mathcal{G}_\Delta(X)$ and $2^{s'}$ is $\Delta$-computable, then $s' \in \mathcal{G}_\Delta(X)$.*

**Proof:** This follows immediately from Lemma 4.3 and Observation 4.4. $\square$

Note that Corollary 4.5 implies that $\mathcal{G}_\Delta(X)$ is a dense subset of the interval $[\dim_\Delta(X), \infty)$.

The following consequence of Lemma 4.3 says that supergales can be used to establish upper bounds on dimension.

**Corollary 4.6.** *Let $X \subseteq \mathbf{C}$ and $s \in [0,\infty)$. If there is a $\Delta$-computable $s$-supergale $d$ such that $X \subseteq S^\infty[d]$, then $\dim_\Delta(X) \leq s$.*

**Proof:** Assume the hypothesis, and let $s' \geq s$ be arbitrary such that $2^{s'}$ is $\Delta$-computable. By Observation 4.4, $d$ is a $\Delta$-computable $s'$-supergale with $X \subseteq S^\infty[d]$. It follows by Lemma 4.3 that $s' \in \mathcal{G}_\Delta(X)$, whence $\dim_\Delta(X) \leq s'$. Since this holds for all $s' \geq s$ with $2^{s'}$ $\Delta$-computable, it follows that $\dim_\Delta(X) \leq s$. $\square$

**Lemma 4.7.** (Exact Computation Lemma). *If $d$ is a $\Delta$-computable $s$-supergale and $2^s$ is rational, then there is an exactly $\Delta$-computable $s$-gale $\tilde{d}$ such that $S^\infty[d] \subseteq S^\infty[\tilde{d}]$.*

**Proof:** Assume the hypothesis. If $s = 0$, then $S^\infty[d] = \emptyset$ and the conclusion holds trivially, so assume that $s > 0$. By Lemma 4.3, it suffices to show that there is an exactly computable $s$-supergale $\tilde{d}$ such that $S^\infty[d] \subseteq S^\infty[\tilde{d}]$.

Since $d$ is $\Delta$-computable, there is an exactly $\Delta$-computable function $\hat{d} : \{0,1\}^* \times \mathbb{N} \longrightarrow \mathbb{Q} \cap [0,\infty)$ such that for all $w \in \{0,1\}^*$ and $r \in \mathbb{N}$, $|\hat{d}(w,r) - d(w)| \leq 2^{-r}$. Let

$$a = 1 + \left\lceil \log \frac{1}{1 - 2^{-s}} \right\rceil,$$

so that $2^{1-a} \leq 1 - 2^{-s}$, and define

$$\tilde{d} : \{0,1\}^* \longrightarrow \mathbb{Q} \cap [0,\infty)$$

$$\tilde{d}(w) = \hat{d}(w, |w| + a) + 2^{-|w|} .$$

It is clear that $\tilde{d}$ is exactly $\Delta$-computable. Also, for all $w \in \{0,1\}^*$,

$$\begin{aligned}
&\tilde{d}(w) - 2^{-s}[\tilde{d}(w0) + \tilde{d}(w1)] \\
&\geq d(w) - 2^{-(|w|+a)} + 2^{-|w|} - 2^{-s}\left[d(w0) + d(w1) + 2\left[2^{-(|w|+a+1)} + 2^{-(|w|+1)}\right]\right] \\
&= d(w) - 2^{-s}[d(w0) + d(w1)] + 2^{-|w|}[1 - 2^{-s} - (1 + 2^{-s})2^{-a}] \\
&\geq 2^{-|w|}[1 - 2^{-s} - 2^{1-a}] \\
&\geq 0 ,
\end{aligned}$$

so $\tilde{d}$ is an $s$-supergale. Finally, for all $w \in \{0,1\}^*$,

$$\tilde{d}(w) - d(w) \geq 2^{-|w|} - 2^{-(|w|+a)} > 0 ,$$

so $S^\infty[d] \subseteq S^\infty[\tilde{d}]$. □

**Lemma 4.8.** *For all $X, Y \subseteq \mathbf{C}$,*

$$\dim_\Delta(X \cup Y) = \max\{\dim_\Delta(X), \dim_\Delta(Y)\}$$

*and*

$$\dim(X \cup Y \mid R(\Delta)) = \max\{\dim(X \mid R(\Delta)), \dim(Y \mid R(\Delta)).$$

***Proof:*** The second identity follows from the first, so by Observation 4.1 it suffices to show that

$$\dim_\Delta(X \cup Y) \leq \max\{\dim_\Delta(X), \dim_\Delta(Y)\} .$$

For this, choose an arbitrary $s > \max\{\dim_\Delta(X), \dim_\Delta(Y)\}$ such that $2^s$ is $\Delta$-computable. By Corollary 4.5, $s \in \mathcal{G}_\Delta(X) \cap \mathcal{G}_\Delta(Y)$, so there exist $\Delta$-computable $s$-gales $d_X$ and $d_Y$ such that $X \subseteq S^\infty[d_X]$ and $Y \subseteq S^\infty[d_Y]$. Let $d = d_X + d_Y$. Then $d$ is clearly $\Delta$-computable, and $d$ is an $s$-gale by Observation 3.5. It is clear that $X \cup Y \subseteq S^\infty[d]$, whence $s \in \mathcal{G}_\Delta(X \cup Y)$. It follows that $\dim_\Delta(X \cup Y) < s$. Since $s$ is arbitrary here, we have shown that $\dim_\Delta(X \cup Y) \leq \max\{\dim_\Delta(X), \dim_\Delta(Y)\}$. □

Lemma 4.8 has the following immediate consequence.

**Corollary 4.9.** *For all $X \in \mathbf{C}$ and $n \in \mathbb{N}$,*

$$\dim_\Delta(X) = \max_{w \in \{0,1\}^n} \dim_\Delta(X \cap \mathbf{C}_w)$$

*and*

$$\dim(X \mid R(\Delta)) = \max_{w \in \{0,1\}^n} \dim(X \cap \mathbf{C}_w \mid R(\Delta)) .$$

A set $X \subseteq \mathbf{C}$ is *closed under finite variations* if for every $A \in X$ and every finite set $D \subseteq \{0,1\}^*$, we have $A \triangle D \in X$, where $A \triangle D = (A - D) \cup (D - A)$ is the symmetric difference of $A$ and $D$. A set $X$ with this property is called a *tail set*.

**Corollary 4.10.** *If $X \subseteq \mathbf{C}$ is a tail set, then for all $w \in \{0,1\}^*$,*

$$\dim_\Delta(X \cap \mathbf{C}_w) = \dim_\Delta(X)$$

*and*

$$\dim(X \cap \mathbf{C}_w \mid R(\Delta)) = \dim(X \mid R(\Delta)) .$$

Lemma 4.8 can be extended to countable unions, provided that these unions are sufficiently constructive.

**Definition.** *Let $X, X_0, X_1, X_2, \ldots \subseteq \mathbf{C}$.*

1. *$X$ is a $\Delta$-union of the $\Delta$-dimensioned sets $X_0, X_1, X_2, \ldots$ if $X = \bigcup_{k=0}^{\infty} X_k$ and for each $s > \sup_{k \in \mathbb{N}} \dim_{\Delta}(X_k)$ with $2^s$ rational, there is a function $d : \mathbb{N} \times \{0,1\}^* \to [0, \infty)$ with the following properties.*

    *(i) $d$ is $\Delta$-computable.*
    *(ii) For each $k \in \mathbb{N}$, if we write $d_k(w) = d(k, w)$, then the function $d_k$ is an $s$-gale.*
    *(iii) For each $k \in \mathbb{N}$, $X_k \subseteq S^{\infty}[d_k]$.*

2. *$X$ is a $\Delta$-union of the sets $X_0, X_1, X_2, \ldots$ dimensioned in $R(\Delta)$ if $X = \bigcup_{k=0}^{\infty} X_k$ and $X \cap R(\Delta)$ is an $\Delta$-union of the $\Delta$-dimensional sets $X_0 \cap R(\Delta), X_1 \cap R(\Delta), X_2 \cap R(\Delta), \ldots$ .*

**Lemma 4.11.** *Let $X, X_0, X_1, X_2, \ldots \subseteq \mathbf{C}$.*

1. *If $X$ is a $\Delta$-union of the $\Delta$-dimensioned sets $X_0, X_1, X_2, \ldots,$ then*

$$\dim_{\Delta}(X) = \sup_{k \in \mathbb{N}} \dim_{\Delta}(X_k) \ .$$

2. *If $X$ is a $\Delta$-union of the sets $X_0, X_1, X_2, \ldots$ dimensioned in $R(\Delta)$, then*

$$\dim(X \mid R(\Delta)) = \sup_{k \in \mathbb{N}} \dim(X_k \mid R(\Delta)) \ .$$

*Proof:* It suffices to prove 1, since 2 follows immediately from 1. Assume the hypothesis of 1, and let $s > \sup_{k \in \mathbb{N}} \dim_{\Delta}(X_K)$ be arbitrary with $2^s$ rational and $s < 2$. By Observation 4.1, it suffices to show that $\dim_{\Delta}(X) \leq s$.

Since $X$ is a union of the $\Delta$-dimensioned sets $X_0, X_1, X_2, \ldots$, there is a $\Delta$-computable function $d : \mathbb{N} \times \{0,1\}^* \longrightarrow [0, \infty)$ such that each $d_k$ is an $s$-gale with $X_k \subseteq S^{\infty}[d_k]$. Without loss of generality (modifying $d$ if necessary), we can assume that each $d_k(\lambda) \leq 1$.

Let $\tilde{d} = \sum_{k=0}^{\infty} 2^{-k} d_k$. By Observation 3.5, $\tilde{d}$ is an $s$-gale. Since $d$ is $\Delta$-computable, there is a function $\hat{d} : \mathbb{N} \times \mathbb{N} \times \{0,1\}^* \longrightarrow \mathbb{Q} \cap [0, \infty)$ such that $\hat{d} \in \Delta$ and for all $r, k \in \mathbb{N}$ and $w \in \{0,1\}^*$, $|\hat{d}(r, k, w) - d(k, w)| \leq 2^{-r}$. Define

$$\hat{\tilde{d}} : \mathbb{N} \times \{0,1\}^* \longrightarrow \mathbb{Q} \cap [0, \infty)$$

$$\hat{\tilde{d}}(r, w) = \sum_{k=0}^{r+2|w|+1} 2^{-k} \hat{d}(r+2, k, w) \ .$$

Then $\hat{\tilde{d}} \in \Delta$ and for all $r \in \mathbb{N}$ and $w \in \{0,1\}^*$,

$$|\hat{\tilde{d}}(r, w) - \tilde{d}(w)| \leq |\tilde{d}(w) - a| + |a - \hat{\tilde{d}}(w)| \ ,$$

where $a = \sum_{k=0}^{r+2|w|+1} 2^{-k} d_k(w)$. By Corollary 3.4,

$$\begin{aligned}
|\tilde{d}(w) - a| &= \sum_{k=r+2|w|+2}^{\infty} 2^{-k} d_k(w) \\
&\leq \sum_{k=r+2|w|+2}^{\infty} 2^{-k} 2^{s|w|} d_k(\lambda) \\
&\leq \sum_{k=r+2|w|+2}^{\infty} 2^{2|w|-k} \\
&= 2^{-(r+1)} .
\end{aligned}$$

Also,

$$\begin{aligned}
|a - \hat{\tilde{d}}(w)| \\
&\leq \sum_{k=0}^{r+2|w|+1} 2^{-k} |\tilde{d}(r+2, k, w) - d(k, w)| \\
&\leq \sum_{k=0}^{\infty} 2^{-(k+r+2)} \\
&= 2^{-(r+1)}
\end{aligned}$$

It follows that for all $r \in \mathbb{N}$ and $w \in \{0,1\}^*$,

$$|\hat{\tilde{d}}(r, w) - \tilde{d}| \leq 2^{-r} ,$$

whence $\hat{\tilde{d}}$ testifies that $\tilde{d}$ is $\Delta$-computable. It is clear that $X = \bigcup_{k=0}^{\infty} X_k \subseteq \bigcup_{k=0}^{\infty} S^{\infty}[d_k] \subseteq S^{\infty}[\tilde{d}]$, so it follows that $\dim_\Delta(X) \leq s$. □

We now note that finite sets of languages have resource-bounded dimension 0, provided that the languages themselves do not exceed the resource bound.

**Lemma 4.12.** *If $X \subseteq R(\Delta)$ is finite, then $\dim(X \mid R(\Delta)) = \dim_\Delta(X) = 0$.*

Lemma 4.12 can be extended to subsets of $R(\Delta)$ that are "countable within the resource bound $\Delta$" in the following sense. A set $X \subseteq \mathbf{C}$ is $\Delta$-*countable* if there is a function $\delta : \mathbb{N} \times \{0,1\}^* \to \{0,1\}^*$ with the following properties.

(i) $\delta \in \Delta$.

(ii) For each $k \in \mathbb{N}$, if we write $\delta_k(w) = \delta(k, w)$, then the function $\delta_k$ is a constructor.

(iii) $X = \{R(\delta_k) \mid k \in \mathbb{N}\}$ .

**Proof:** By Lemma 4.8, it suffices to prove this for singleton sets $X \subseteq R(\Delta)$, so assume that $X = \{A\}$, where $A \in R(\Delta)$. Let $s > 0$ be arbitrarily small with $2^s$ rational. It suffices to show that $\dim_\Delta(X) \leq s$. Define

$$d : \{0,1\}^* \longrightarrow [0, \infty)$$

$$d(w) = \begin{cases} 2^{s|w|} & \text{if } w \sqsubseteq A \\ 0 & \text{if } w \not\sqsubseteq A . \end{cases}$$

Then $d \in \Delta$ (because $A \in R(\Delta)$ and $2^s$ is rational) and it is clear that $d$ is an $s$-gale that succeeds on $A$. Thus $\dim_\Delta(X) \leq s$. □

It is clear that if $X$ is $\Delta$-countable, then $X \subseteq R(\Delta)$. It was shown in [9] that every $\Delta$-countable set has $\Delta$-measure 0. We now show that more is true.

**Lemma 4.13.** *If $X \subseteq \mathbf{C}$ is $\Delta$-countable, then $\dim(X \mid R(\Delta)) = \dim_\Delta(X) = 0$.*

**Proof:** Let the function $\delta : \mathbb{N} \times \{0,1\}^* \longrightarrow \{0,1\}^*$ testify that $X$ is $\Delta$-countable. By Lemmas 4.11 and 4.12, it suffices to show that $X$ is a $\Delta$-union of the $\Delta$-dimension 0 singleton sets $\{R(\delta_0)\}, \{R(\delta_1)\}, \{R(\delta_2)\}, \ldots$ . For this, let $s > 0$ with $2^s$ rational, and define

$$d : \mathbb{N} \times \{0,1\}^* \longrightarrow [0, \infty)$$

$$d(k, w) = \begin{cases} 2^{s|w|} & \text{if } w \sqsubseteq R(\delta_k) \\ 0 & \text{if } w \not\sqsubseteq R(\delta_k) \end{cases}.$$

Then $d$ clearly has the required properties. □

If $\Delta = $ all, then Lemma 4.13 is the well-known fact that every countable set has Hausdorff dimension 0. For smaller resource bounds $\Delta$, Lemma 4.13 has consequences of the following sort.

**Corollary 4.14.** *For every $k \in \mathbb{N}$,*

$$\dim(\mathrm{DTIME}(2^{kn}) \mid \mathrm{E}) = 0$$

*and*

$$\dim(\mathrm{DTIME}(2^{n^k}) \mid \mathrm{E}_2) = 0 .$$

Analogous results hold in ESPACE, REC, etc. As noted above, every countable class of languages has Hausdorff dimension 0. In contrast, we now show that, even for countable resource bounds $\Delta$, the $\Delta$-dimension of $R(\Delta)$ is 1. This result, which is analogous to the Measure Conservation Theorem of [9], endows the classes $R(\Delta)$ with internal dimensional structure. (In fact, by Observation 4.1, this result follows from the Measure Conservation Theorem, but we prove it directly here.)

**Theorem 4.15.** $\dim(R(\Delta) \mid R(\Delta)) = \dim_\Delta(R(\Delta)) = 1$.

**Proof:** By definition, $\dim(R(\Delta) \mid R(\Delta)) = \dim_\Delta(R(\Delta))$ and by Observation 4.1, $\dim_\Delta(R(\Delta)) \leq 1$, so it suffices to show that $\dim_\Delta(R(\Delta)) \geq 1$. For this, fix an arbitrary $s \in [0, 1)$ such that $2^s$ is rational. By Corollary 4.5, it suffices to show that $s \notin \mathcal{G}_\Delta(R(\Delta))$. For this, let $d \in \Delta$ be an exact $s$-gale. By the Exact Computation Lemma, it suffices to show that $R(\Delta) \not\subseteq S^\infty[d]$. We do this by defining a constructor $\delta \in \Delta$ such that $R(\delta) \notin S^\infty[d]$.

For each $w \in \{0,1\}^*$, let

$$\delta(w) = w[\![d(w0) > d(w1)]\!] .$$

Then $\delta$ is a constructor, and it is clear that $\delta \in \Delta$ (because $d \in \Delta$). The definition of $\delta$ ensures that for all $w \in \{0,1\}^*$,

$$d(\delta(w)) = \min\{d(w0), d(w1)\} .$$

It follows by Corollary 3.3 that for all $w \in \{0,1\}^*$, $d(\delta(w)) \leq 2^{s-1}d(w)$. By induction, this implies that for all $n \in \mathbb{N}$, $d(\delta^n(\lambda)) \leq 2^{(s-1)n}d(\lambda)$. Since $s \in [0, 1)$, we then have

$$\lim_{n \to \infty} d(R(\delta)[0 \ldots n-1]) = \lim_{n \to \infty} d(\delta^n(\lambda)) = 0 ,$$

whence $R(\delta) \notin S^\infty[d]$. □

We now give an example in which we calculate the resource-bounded dimension of a simple set of languages.

**Proposition 4.16.** *Given $l \in \mathbb{Z}^+$ and $\emptyset \neq S \subseteq \{0,1\}^l$, let*

$$X = \{A \in \mathbf{C} \mid (\forall k) A[kl..(k+1)l - 1] \in S\} \ .$$

*Then*

$$\dim(X \mid \mathrm{E}) = \dim_\mathrm{p}(X) = \frac{\log |S|}{l} \ .$$

**Proof:** Since $\dim(X \mid \mathrm{E}) \leq \dim_\mathrm{p}(X)$, it suffices to show that $\dim_\mathrm{p}(X) \leq \frac{\log |S|}{l}$ and $\dim(X \mid \mathrm{E}) \geq \frac{\log |S|}{l}$.

To see that $\dim_\mathrm{p}(X) \leq \frac{\log |S|}{l}$, let $s > \frac{\log |S|}{l}$ be such that $2^s$ is rational. Define a function $d : \{0,1\}^* \longrightarrow [0, \infty)$ inductively as follows. Let $d(\lambda) = 1$. If $d(w)$ has been defined, where $|w|$ is a multiple of $l$, and if $0 < |u| \leq l$, then let $d(wu) = 2^{s|u|}\rho(u)d(w)$, where

$$\rho(u) = \frac{|\{v \in S \mid u \sqsubseteq v\}|}{|S|} \ .$$

It is clear that $d$ is exactly p-computable, and it is routine to verify that $d$ is an $s$-gale. The definition of $d$ implies that if $|w|$ is a multiple of $l$ and $u \in S$, then

$$d(wu) = 2^{sl}\frac{1}{|S|}d(w) = 2^\varepsilon d(w) \ ,$$

where $\varepsilon = sl - \log |S| > 0$ by our choice of $s$. It follows inductively that if $A \in X$, then for all $k \in \mathbb{N}$,

$$d(A[0 \ldots kl - 1]) = 2^{\varepsilon k} \ ,$$

whence $A \in S^\infty[d]$. Thus $X \subseteq S^\infty[d]$. We have now established that $\dim_\mathrm{p}(X) \leq s$. Since this holds for a dense set of $s > \frac{\log |S|}{l}$, it follows that $\dim_\mathrm{p}(X) \leq \frac{\log |S|}{l}$.

To see that $\dim(X \mid \mathrm{E}) \geq \frac{\log |S|}{l}$, let $0 \leq s < \frac{\log |S|}{l}$ be such that $2^s$ is rational, and let $d$ be an exactly p-computable $s$-gale. By the Exact Computation Lemma, it suffices to show that $X \cap \mathrm{E} \not\subseteq S^\infty[d]$. Define a constructor $\delta$ as follows. If $|w|$ is a multiple of $l$, then $\delta(w) = wu$, where $u$ is the lexicographically first element of $S$ such that for all $v \in S$, $d(wu) \leq d(wv)$. If $|w|$ is not a multiple of $l$, then $\delta(w) = w0$. Since $\delta$ is exactly p-computable, it is clear that $\delta \in \mathrm{p}$. It is then easy to see that $R(\delta) \in X \cap R(\mathrm{p}) = X \cap \mathrm{E}$. We finish the proof by showing that $R(\delta) \notin S^\infty[d]$.

For any string $w$, Corollary 3.3 (with $\alpha = \frac{2^l}{|S|}$) tells us that there are fewer than $|S|$ strings $u \in \{0,1\}^l$ for which $d(wu) > \frac{2^{sl}}{|S|}d(w)$. This implies that for all $w$ such that $|w|$ is a multiple of $l$, $d(\delta(w)) \leq \frac{2^{sl}}{|S|}d(w)$. Thus for all $k \in \mathbb{N}$,

$$d(R(\delta)[0 \ldots kl - 1]) \leq \left(\frac{2^{sl}}{|S|}\right)^k d(\lambda) = 2^{-\varepsilon k}d(\lambda) \ ,$$

where $\varepsilon = \log |S| - sl > 0$. It follows by Corollary 3.4 that for all $n \in \mathbb{N}$,

$$d(R(\delta)[0 \ldots n - 1]) \leq 2^{sl-\varepsilon \lfloor \frac{n}{l} \rfloor}d(\lambda)$$

whence $R(\delta) \notin S^{\infty}[d]$.  □

We conclude this section by mentioning some relevant earlier work relating martingales and supermartingales to computable dimension. Schnorr [14, 15] defined a martingale $d$ to have *exponential order* on a sequence (equivalently, language) $S$ if

$$\limsup_{n \to \infty} \frac{\log d(S[0..n-1])}{n} > 0 \tag{4.1}$$

and proved that no computable martingale can have exponential order on a Church-stochastic sequence. Terwijn [18] has noted that (4.1) is equivalent to the existence of an $s < 1$ for which the $s$-gale $d^{(s)}(w) = 2^{(s-1)|w|}d(w)$ succeeds on $S$. Thus Schnorr's result says that $\dim_{\text{comp}}(\{S\}) = 1$ for every Church-stochastic sequence $S$.

Ryabko [13] and Staiger [17] defined the *exponent of increase* $\lambda_d(S)$ of a martingale $d$ on a sequence $S$ to be the left-hand side of (4.1). (We are using Staiger's notation here.) Both papers paid particular attention to the quantity

$$\lambda(S) = \sup\{\lambda_d(S) | d \text{ is a computable martingale}\}. \tag{4.2}$$

By Terwijn's above-mentioned observation, $1 - \lambda(S)$ is precisely $\dim_{\text{comp}}(\{S\})$. The reader is referred to these papers for interesting results relating $\lambda(S)$ – and hence computable dimension – to Kolmogorov complexity and Hausdorff dimension.

## 5 Dimension and Frequency in Exponential Time

In this section we show that for each $0 \le s \le 1$ there is a natural set $X$ that has dimension $s$ in each of the exponential time complexity classes E and $E_2$. This set $X$ consists of those languages that asymptotically contain at most $\alpha$ of all strings, where $0 \le \alpha \le \frac{1}{2}$ and $\mathcal{H}(\alpha) = s$. We now define this set more completely.

For each nonempty string $w \in \{0,1\}^+$, let

$$\text{freq}(w) = \frac{\#(1, w)}{|w|} ,$$

where $\#(1, w)$ is the number of 1's in $w$. For each $A \in \mathbf{C}$ and $n \in \mathbb{Z}^+$, let

$$\text{freq}_A(n) = \text{freq}(A[0..n-1]) .$$

That is, $\text{freq}_A(n)$ is the fraction of the first $n$ strings in $\{0,1\}^*$ that are elements of $A$. For $\alpha \in [0,1]$, define the sets

$$\text{FREQ}(\alpha) = \{A \in \mathbf{C} \mid \lim_{n \to \infty} \text{freq}_A(n) = \alpha\} ,$$

$$\text{FREQ}(\le \alpha) = \{A \in \mathbf{C} \mid \liminf_{n \to \infty} \text{freq}_A(n) \le \alpha\} .$$

The set $\text{FREQ}(\le \alpha)$ is the set $X$ promised in the preceding paragraph.

Our results use the binary entropy function

$$\mathcal{H} : [0,1] \longrightarrow [0,1]$$

$$\mathcal{H}(\alpha) = \alpha \log \frac{1}{\alpha} + (1-\alpha) \log \frac{1}{1-\alpha} .$$

(The values $\mathcal{H}(0)$ and $\mathcal{H}(1)$ are both 0, so that $\mathcal{H}$ is continuous on $[0, 1]$.) Our proofs use the "weighted entropy" function
$$h : (0,1) \times (0,1) \longrightarrow \mathbb{R}$$
$$h(x,y) = x \log \frac{1}{y} + (1-x) \log \frac{1}{1-y}.$$

For fixed $x$, $h(x,y)$ takes its minimum value $\mathcal{H}(x)$ at $y = x$ and strictly increases as $y$ moves away from $x$.

We first show that $\mathcal{H}(\alpha)$ is an upper bound on the p-dimension of $\mathrm{FREQ}(\leq \alpha)$.

**Lemma 5.1.** *For all $\alpha \in [0, \frac{1}{2}]$, $\dim_\mathrm{p}(\mathrm{FREQ}(\leq \alpha)) \leq \mathcal{H}(\alpha)$.*

***Proof:*** Let $0 \leq \alpha \leq \frac{1}{2}$. Let $s > \mathcal{H}(\alpha)$ be such that $2^s$ is rational, and let $\varepsilon = \frac{s - \mathcal{H}(\alpha)}{2}$. Fix $\delta > 0$ such that for all $x, y \in [\alpha - \delta, \alpha + \delta] \cap (0, 1)$, $|h(x,y) - \mathcal{H}(\alpha)| < \varepsilon$, and let $y \in [\alpha - \delta, \alpha + \delta] \cap (0, \frac{1}{2}]$ be a rational number. Define
$$d : \{0,1\}^* \longrightarrow \mathbb{Q} \cap [0, \infty)$$
$$d(\lambda) = 1$$
$$d(w0) = (1-y)2^s d(w)$$
$$d(w1) = y 2^s d(w) \ .$$

It is clear that $d$ is an exactly p-computable $s$-gale.

To see that $\mathrm{FREQ}(\leq \alpha) \subseteq S^\infty[d]$, let $A \in \mathrm{FREQ}(\leq \alpha)$. Then there exists an infinite set $J \subseteq \mathbb{N}$ such that for all $n \in J$, $\mathrm{freq}_A(n) \leq \alpha + \delta$. It follows that for all $n \in J$, if we write $w_n = A[0 \ldots n-1]$, then
$$\begin{aligned}
d(w_n) &= y^{\#(1,w_n)}(1-y)^{\#(0,w_n)} 2^{sn} \\
&= \left[ y^{\mathrm{freq}_A(n)}(1-y)^{1-\mathrm{freq}_A(n)} 2^s \right]^n \\
&\geq \left[ y^{\alpha+\delta}(1-y)^{1-(\alpha+\delta)} 2^s \right]^n \\
&= \left[ 2^{s-h(\alpha+\delta,y)} \right]^n \\
&\geq \left[ 2^{s-\mathcal{H}(\alpha)-\varepsilon} \right]^n \\
&= 2^{\varepsilon n} \ .
\end{aligned}$$

Thus $A \in S^\infty[d]$.

We have now shown that $\dim_\mathrm{p}(\mathrm{FREQ}(\leq \alpha)) \leq s$. Since this holds for all $s > \mathcal{H}(\alpha)$ such that $2^s$ is rational, it follows that $\dim_\mathrm{p}(\mathrm{FREQ}(\leq \alpha)) \leq \mathcal{H}(\alpha)$. □

We next show that $\mathcal{H}(\alpha)$ is a lower bound on the dimension of $\mathrm{FREQ}(\alpha)$ in $R(\Delta)$. In general, this does not hold for arbitrary $\alpha \in [0,1]$, since if $R(\Delta)$ is countable there can be only countably many $\alpha$ for which $\mathrm{FREQ}(\alpha) \cap R(\Delta) \neq \emptyset$. However, it does hold for all $\Delta$-computable real numbers $\alpha \in [0, 1]$.

**Lemma 5.2.** *For all $\Delta$-computable $\alpha \in [0,1]$, $\dim(\mathrm{FREQ}(\alpha) \mid R(\Delta)) \geq \mathcal{H}(\alpha)$.*

***Proof:*** The inequality holds trivially for $\alpha \in \{0, 1\}$, so assume that $\alpha \in (0, 1)$ is $\Delta$-computable. Fix $s \in (0, \mathcal{H}(\alpha))$ such $2^s$ is rational, and let $d$ be an exactly $\Delta$-computable $s$-gale. By the Exact Computation Lemma, it suffices to show that $\text{FREQ}(\alpha) \cap R(\Delta) \not\subseteq S^\infty[d]$.

Although the details are a bit involved, the idea of the proof is simple. We want to define a constructor $\delta \in \Delta$ such that $R(\delta) \in \text{FREQ}(\alpha) - S^\infty[d]$. Given a string $w$ of length $n$, $\delta$ computes a rational approximation $\frac{k(n)}{m(n)}$ of $\alpha$ and extends $w$ by a string $u$ of length $m(n)$ containing exactly $k(n)$ 1's and having the property that $d$ makes no progress along $u$. Such a string $u$ exists because $s < \mathcal{H}\left(\frac{k(n)}{m(n)}\right)$, and it can be found within the resource bound $\Delta$ because $m(n)$ is logarithmic in $n$. On the other hand, $m(n) \to \infty$ as $n \to \infty$, so $R(\delta) \in \text{FREQ}(\alpha)$, and $d$ makes no progress along $R(\delta)$, so $R(\delta) \notin S^\infty[d]$. We now develop this idea.

Since $\alpha$ is $\Delta$-computable, there is a function $\hat{\alpha} : \mathbb{N} \to \mathbb{Q} \cap (0, 1)$ such that $\hat{\alpha} \in \Delta$ and for all $r \in \mathbb{N}$, $|\hat{\alpha}(r) - \alpha| \leq 2^{-r}$. Fix $\epsilon > 0$ such that

$$\epsilon \leq \alpha, \quad \epsilon \leq 1 - \alpha, \tag{5.1}$$

and for all $x \in [0, 1]$,

$$|x - \alpha| < \epsilon \Rightarrow |\mathcal{H}(x) - \mathcal{H}(\alpha)| < \frac{\mathcal{H}(\alpha) - s}{2}. \tag{5.2}$$

Choose a positive integer $c$ satisfying the conditions

$$c > 2^{1+\frac{2}{\epsilon}}, \tag{5.3}$$

$$c > 1 + \log \frac{1}{\epsilon}, \tag{5.4}$$

$$\frac{\log c}{\log \log c} > \frac{4}{\mathcal{H}(\alpha) - s}. \tag{5.5}$$

For each $n \in \mathbb{N}$, let

$$m(n) = \lfloor \log(n + c) \rfloor,$$
$$k(n) = \lfloor \hat{\alpha}(m(n) + c) m(n) \rfloor.$$

We first show that $\frac{k(n)}{m(n)}$ is a useful approximation of $\alpha$. By (5.3),

$$m(n) \geq \lfloor \log c \rfloor \geq \frac{2}{\epsilon} \tag{5.6}$$

for all $n \in \mathbb{N}$. In particular, $m(n)$ is always positive. By (5.4) and (5.1),

$$\begin{aligned} \hat{\alpha}(m(n) + c) &\geq \alpha - 2^{-c} \geq \alpha - 2^{-1+\log \epsilon} \\ &= \alpha - \frac{\epsilon}{2} \geq \frac{\alpha}{2} \end{aligned} \tag{5.7}$$

for all $n \in \mathbb{N}$. It follows from (5.6), (5.7), and (5.1) that

$$k(n) \geq \left\lfloor \frac{2}{\epsilon} \frac{\alpha}{2} \right\rfloor > 0 \tag{5.8}$$

for all $n \in \mathbb{N}$. Also by (5.4) and (5.1),

$$\begin{aligned} \hat{\alpha}(m(n) + c) &\leq \alpha + 2^{-c} < \alpha + 2^{-1+\log \epsilon} \\ &\leq \alpha + 2^{-1+\log(1-\alpha)} = \frac{1 + \alpha}{2} < 1, \end{aligned}$$

so
$$k(n) = \lfloor \hat{\alpha}(m(n)+c)m(n) \rfloor < m(n) \tag{5.9}$$

for all $n \in \mathbb{N}$. By (5.8) and (5.9), we now have

$$0 < \frac{k(n)}{m(n)} < 1 \tag{5.10}$$

for all $n \in \mathbb{N}$. In fact,

$$\frac{k(n)}{m(n)} = \frac{\lfloor \hat{\alpha}(m(n)+c)m(n) \rfloor}{m(n)},$$

so

$$\hat{\alpha}(m(n)+c) - \frac{1}{m(n)} < \frac{k(n)}{m(n)} \leq \hat{\alpha}(m(n)+c),$$

so

$$\left| \frac{k(n)}{m(n)} - \alpha \right| \leq \frac{1}{m(n)} + 2^{-(m(n)+c)} \tag{5.11}$$

for all $n \in \mathbb{N}$. It follows by (5.4) and (5.6) that

$$\left| \frac{k(n)}{m(n)} - \alpha \right| \leq \epsilon,$$

whence by (5.2),

$$\left| \mathcal{H}\left(\frac{k(n)}{m(n)}\right) - \mathcal{H}(\alpha) \right| < \frac{\mathcal{H}(\alpha) - s}{2} \tag{5.12}$$

for all $n \in \mathbb{N}$. This is the first sense in which $\frac{k(n)}{m(n)}$ is a useful approximation of $\alpha$.

For each $n \in \mathbb{N}$, define the set

$$B_n = \left\{ u \in \{0,1\}^{m(n)} \,\middle|\, \#(1,u) = k(n) \right\}.$$

Using (5.10) and the well-known approximation $e\left(\frac{N}{e}\right)^N < N! < eN\left(\frac{N}{e}\right)^N$, valid for all $N \geq 1$, it is easy to see that

$$\begin{aligned} |B_n| &= \binom{m(n)}{k(n)} > \frac{1}{ek(n)(m(n)-k(n))} 2^{m(n)\mathcal{H}\left(\frac{k(n)}{m(n)}\right)} \\ &\geq \frac{4}{em(n)^2} 2^{m(n)\mathcal{H}\left(\frac{k(n)}{m(n)}\right)} \\ &> 2^{m(n)\mathcal{H}\left(\frac{k(n)}{m(n)}\right) - 2\log m(n)} \end{aligned}$$

for all $n \in \mathbb{N}$. It follows by (5.12) that

$$\begin{aligned} |B_n| &> 2^{m(n)\left[\mathcal{H}(\alpha) - \frac{\mathcal{H}(\alpha)-s}{2}\right] - 2\log m(n)} \\ &= 2^{m(n)\frac{\mathcal{H}(\alpha)+s}{2} - 2\log m(n)} \end{aligned}$$

for all $n \in \mathbb{N}$. Now (5.5) and the monotonicity of $\frac{\log x}{\log \log x}$ tell us that

$$m(n)\frac{\mathcal{H}(\alpha) - s}{2} \geq 2\log m(n),$$

so it follows that

$$|B_n| > 2^{m(n)\left[\frac{\mathcal{H}(\alpha)+s}{2} - \frac{\mathcal{H}(\alpha-s)}{2}\right]} = 2^{sm(n)} \tag{5.13}$$

for all $n \in \mathbb{N}$.

Corollary 3.3 tells us that for each $w \in \{0,1\}^*$ there are fewer than $2^{sm(n)}$ strings $u \in \{0,1\}^{m(n)}$ for which $\max_{v \sqsubseteq u} d(wv) > d(w)$. It follows by (5.13) that the function

$$\delta : \{0,1\}^* \to \{0,1\}^*$$

$$\delta(w) = wu, \quad \text{where } u \text{ is the lexicographically}$$
$$\text{first element of } B_{|w|} \text{ such that}$$
$$d(wv) \leq d(w) \text{ for all } v \sqsubseteq u$$

is a well-defined constructor. Also, since $d \in \Delta$ and $|\{0,1\}^{m(|w|)}| \leq |w| + c$, it is clear that $\delta \in \Delta$. To conclude the proof, then, it suffices to show that $R(\delta) \in \mathrm{FREQ}(\alpha) - S^\infty[d]$.

To see that $R(\delta) \in \mathrm{FREQ}(\alpha)$, define a sequence $n_0, n_1, \ldots$ by the recursion

$$n_0 = 0, \quad n_{i+1} = n_i + m(n_i),$$

and note that $R(\delta)$ is of the form

$$R(\delta) = u_0 u_1 u_2 \cdots$$

where $|u_i| = m(n_i)$ and $\mathrm{freq}(u_i) = \frac{k(n_i)}{m(n_i)}$ for all $i \in \mathbb{N}$. Also, by (5.11) we have

$$\left|\frac{k(n)}{m(n)} - \alpha\right| \leq \frac{2}{m(n)}$$

for all $n \in \mathbb{N}$. Since $m(n) \to \infty$ as $n \to \infty$, it follows that

$$\lim_{i \to \infty} \mathrm{freq}(u_i) = \lim_{i \to \infty} \frac{k(n_i)}{m(n_i)} = \alpha.$$

(This is the second sense in which $\frac{k(n)}{m(n)}$ is a useful approximation of $\alpha$.) Since $|u_i|$ is nondecreasing, this implies that $R(\delta) \in \mathrm{FREQ}(\alpha)$.

The definition of $\delta$ implies that $d(w) \leq d(\lambda)$ for all $w \sqsubseteq R(\delta)$, so $R(\delta) \notin S^\infty[d]$. □

From these two lemmas we get the main result of this section.

**Theorem 5.3.**

1. For all $\Delta$-computable $\alpha \in [0,1]$, $\dim(\mathrm{FREQ}(\alpha) \mid R(\Delta)) = \mathcal{H}(\alpha)$.

2. For all $\alpha \in [0, \frac{1}{2}]$, $\dim(\mathrm{FREQ}(\leq \alpha)|R(\Delta)) = \mathcal{H}(\alpha)$.

***Proof:***

1. This follows immediately from Lemma 5.1, Lemma 5.2, and Observation 4.1.

2. Let $\alpha \in [0, \frac{1}{2}]$. By Lemma 5.1, and Observation 4.1,

$$\begin{aligned} &\dim(\text{FREQ}(\leq \alpha) \mid R(\Delta)) \\ &= \dim_\Delta(\text{FREQ}(\leq \alpha) \cap R(\Delta)) \\ &\leq \dim_\Delta(\text{FREQ}(\leq \alpha)) \\ &\leq \dim_\text{p}(\text{FREQ}(\leq \alpha)) \\ &\leq \mathcal{H}(\alpha). \end{aligned}$$

For the reverse inequality, let $\alpha'$ be an arbitrary rational such that $\alpha' \leq \alpha$. Then by Lemma 5.2 and Observation 4.1,

$$\begin{aligned} &\dim(\text{FREQ}(\leq \alpha) \mid R(\Delta)) \\ &\geq \dim(\text{FREQ}(\alpha') \cap R(\Delta)) \\ &\geq \mathcal{H}(\alpha') . \end{aligned}$$

Since this holds for all rational $\alpha' \leq \alpha$ and $\mathcal{H}$ is continuous, it follows that

$$\dim(\text{FREQ}(\leq \alpha) \mid R(\Delta)) \geq \mathcal{H}(\alpha) .$$

□

The case $\Delta =$ all of Theorem 5.3 says simply that the classical Hausdorff dimensions of $\text{FREQ}(\alpha)$ and $\text{FREQ}(\leq \alpha)$ are both $\mathcal{H}(\alpha)$. This was proven in 1949 by Eggleston[5, 2]. The proof here yields a new proof, using gales, of this classical result. However, it is complexity-theoretic results of the following kind that are of interest in this paper.

**Corollary 5.4.** 1. For all p-computable reals $\alpha \in [0, 1]$,

$$\dim(\text{FREQ}(\alpha) \mid \text{E}) = \dim(\text{FREQ}(\alpha) \mid \text{E}_2) = \mathcal{H}(\alpha) .$$

2. For all $\alpha \in [0, \frac{1}{2}]$,

$$\dim(\text{FREQ}(\leq \alpha) \mid \text{E}) = \dim(\text{FREQ}(\leq \alpha) \mid \text{E}_2) = \mathcal{H}(\alpha) .$$

Since $\mathcal{H}(0) = 0$, $\mathcal{H}(\frac{1}{2}) = 1$, and $\mathcal{H}$ is continuous, part 2 of the corollary implies that for every $s \in [0, 1]$ there exists $\alpha \in [0, \frac{1}{2}]$ such that

$$\dim(\text{FREQ}(\leq \alpha) \mid \text{E}) = \dim(\text{FREQ}(\leq \alpha) \mid \text{E}_2) = s .$$

## 6 Dimension and Circuit Size in Exponential Space

We now examine the dimensions of Boolean circuit-size complexity classes in the complexity class ESPACE.

Our Boolean circuit model and terminology are standard. Details may be found in [9], but our result is not sensitive to minor details of the model. The *circuit-size complexity* of a language $A \subseteq \{0,1\}^*$ is the function $CS_A : \mathbb{N} \longrightarrow \mathbb{N}$, where $CS_A(n)$ is the number of gates in the smallest $n$-input Boolean circuit that decides $A \cap \{0,1\}^n$. For each function $f : \mathbb{N} \longrightarrow \mathbb{N}$, we define the circuit-size complexity class

$$\text{SIZE}(f) = \{A \in \mathbf{C} \mid (\forall^\infty n) CS_A(n) \leq f(n)\} .$$

Shannon [16] showed (essentially) that $\mathrm{SIZE}(\alpha\frac{2^n}{n})$ has measure 0 in **C** for all $\alpha < 1$, and Lutz [9] showed that $\mathrm{SIZE}(\alpha\frac{2^n}{n})$ also has measure 0 in ESPACE for all $\alpha < 1$. We now use resource-bounded dimension to give a quantitative refinement of these results.

Assume that all $n$-input Boolean circuits are enumerated in a canonical order in which all circuits of size (number of gates) $t$ precede all those of size $t+1$ for all $t \in \mathbb{Z}^+$. Call a circuit in this enumeration *novel* if there is not previous circuit in the enumeration that decides the same subset of $\{0,1\}^n$. For each $n \in \mathbb{N}$ and $t \in \mathbb{Z}^+$, let $N(n,t)$ be the number of novel $n$-input Boolean circuits with at most $t$ gates. The following specific bound on $N(n,t)$ was proven in [9], but as noted there, the idea is essentially due to Shannon [16].

**Lemma 6.1.** *For all $n \in \mathbb{N}$ and $t > n$, $N(n,t) \leq (48et)^t$.*

**Lemma 6.2.** *For every real $\alpha \in [0,1]$, $\dim_{\mathrm{pspace}}(\mathrm{SIZE}(\alpha\frac{2^n}{n})) \leq \alpha$.*

**Proof:** Let $\alpha \in (0,1]$, and let $s > \alpha$ be such that $2^s$ is rational. Define $d : \{0,1\}^* \longrightarrow [0,\infty)$ inductively is follows.

(i) Let $d(\lambda) = 1$.

(ii) Assume that $d(w)$ has been defined, where $|w| = 2^n - 1$ for some $n \in \mathbb{N}$. For each $u$ with $0 < |u| \leq 2^n$, define $d(wu) = 2^{s|u|}\rho(u)d(w)$, where

$$\rho(u) = \frac{N(n,t,u)}{N(n,t)}$$

and $N(n,t,u)$ is the number of novel $n$-input Boolean circuits with at most $t$ gates that decide the sets $B \subseteq \{0,1\}^n$ whose first $|u|$ decisions are the bits of $u$. It is easy to check that $d$ is an exactly pspace-computable $s$-gale. The definition of $d$ implies that if $|w| = 2^n - 1$ and $u$ is the characteristic string of a set $B \subseteq \{0,1\}^n$ that can be decided by a circuit with at most $\alpha\frac{2^n}{n}$ gates, then for sufficiently large $n$,

$$\begin{aligned}
d(wu) &= 2^{s2^n}\frac{1}{N(n,\alpha\frac{2^n}{n})}d(w) \\
&= 2^{s2^n - \alpha\frac{2^n}{n}[\log 48e + \log\alpha + n - \log n]}d(w) \\
&= 2^{(s-\alpha)2^n + \alpha\frac{2^n}{n}[\log n - \log 48e - \log \alpha]}d(w) \\
&\geq 2^{(s-\alpha)2^n}d(w)
\end{aligned}$$

Since $s - \alpha > 0$, this implies that $\mathrm{SIZE}(\alpha\frac{2^n}{n}) \subseteq S^\infty[d]$. $\square$

**Lemma 6.3.** *If $0 < \beta < \alpha \leq 1$, then for all sufficiently large $n$ there are at least $2^{\beta 2^n}$ different sets $B \subseteq \{0,1\}^n$ that are decided by Boolean circuits of fewer than $\alpha\frac{2^n}{n}$ gates.*

**Proof:** Let $m = n + \log\beta$. Then there are $2^{2^m} = 2^{\beta 2^n}$ different sets $C \subseteq \{0,1\}^m$. If we let $\varepsilon = \frac{\alpha-\beta}{2\alpha}$, so that $\beta = \alpha(1-2\varepsilon)$, then for all sufficiently large $n$, Lupanov [8] has shown that each of these sets is decided by a circuit of at most $\frac{2^m}{m}(1+\varepsilon)$ gates. Now for sufficiently large $n$,

$$\begin{aligned}
\frac{2^m}{m} &= \frac{\beta 2^n}{n + \log\beta} \\
&= \alpha\frac{2^n}{n}(1-2\varepsilon)\left(\frac{n}{n+\log\beta}\right) \\
&< \alpha\frac{2^n}{n}(1-\varepsilon),
\end{aligned}$$

so
$$\frac{2^m}{m}(1+\varepsilon) < \alpha\frac{2^n}{n}(1-\varepsilon^2) < \alpha\frac{2^n}{n}.$$

Thus, for each $C \subseteq \{0,1\}^m$, if we let $B_C = \{w0^{n-m} \mid w \in C\}$, then $B_C$ is decided by a Boolean circuit of fewer than $\alpha\frac{2^n}{n}$ gates. □

Recall that in section 2 we defined constructors and showed that $R(\text{pspace}) = \text{ESPACE}$.

**Lemma 6.4.** *For every real $\alpha \in [0,1]$, $\dim_H(\text{SIZE}(\alpha\frac{2^n}{n})) \geq \alpha$ and $\dim(\text{SIZE}(\alpha\frac{2^n}{n})|\text{ESPACE}) \geq \alpha$.*

**Proof:** This is clear if $\alpha = 0$, so assume that $\alpha \in (0,1]$. Let $0 < s < \alpha$ with $2^s$ rational, and let $d$ be an arbitrary $s$-gale. Define a constructor $\delta : \{0,1\}^* \to \{0,1\}^*$ as follows. If $|w|$ is not of the form $2^n - 1$, then $\delta(w) = w0$. If $|w|$ is of the form $2^n - 1$ (i.e., $w$ is the characteristic string of a subset of $\{0,1\}^{<n}$), then $\delta(w) = wu$, where $u$ is the first string in $\{0,1\}^{2^n}$ that minimizes $\max_{v \sqsubseteq u} d(wv)$, subject to the constraint that $u$ is the characteristic string of a set $B \subseteq \{0,1\}^n$ that is decided by a Boolean circuit with fewer than $\alpha\frac{2^n}{n}$ gates. It is clear that $R(\delta) \in \text{SIZE}(\alpha\frac{2^n}{n})$.

Fix $\beta$ such that $s < \beta < \alpha$. By Lemma 6.3, for all sufficiently large $n$ there are at least $2^{\beta 2^n}$ different sets $B \subseteq \{0,1\}^n$ that are decided by Boolean circuits of fewer than $\alpha\frac{2^n}{n}$ gates. By Corollary 3.3, for all $w$ such that $|w| = 2^n - 1$, there are fewer than $2^{\beta 2^n}$ strings $u \in \{0,1\}^{2^n}$ such that $\max_{v \sqsubseteq u} d(wv) > 2^{|v|(s-\beta)}d(w)$. Taken together, these last two sentences imply that for all sufficiently large $n$ and $w$ with $|w| = 2^n - 1$,

$$\max_{wv \sqsubseteq \delta(w)} d(wv) \leq 2^{|v|(s-\beta)}d(w) \leq d(w).$$

It follows from this that $R(\delta) \notin S^\infty[d]$.

We now have that $R(\delta) \in \text{SIZE}(\alpha\frac{2^n}{n}) - S^\infty[d]$, whence $\text{SIZE}(\alpha\frac{2^n}{n}) \not\subseteq S^\infty[d]$. Since $d$ is arbitrary here, this shows that $\dim_H(\text{SIZE}(\alpha\frac{2^n}{n})) \geq \alpha$.

Now let $d$ be as above, and assume further that $d$ is exactly pspace-computable. Then the constructor $\delta$ defined above is in pspace, so $R(\delta) \in R(\text{pspace}) = \text{ESPACE}$, so we have $\text{SIZE}(\alpha\frac{2^n}{n}) \cap \text{ESPACE} \not\subseteq S^\infty[d]$. It follows by Lemma 4.7 that $\dim(\text{SIZE}(\alpha\frac{2^n}{n})|\text{ESPACE}) \geq \alpha$. □

**Theorem 6.5.** *For every real $\alpha \in [0,1]$, $\dim(\text{SIZE}(\alpha\frac{2^n}{n}) \mid \text{ESPACE}) = \dim_H(\text{SIZE}(\alpha\frac{2^n}{n})) = \alpha$.*

**Proof:** This follows immediately from Lemma 6.2, Lemma 6.4, and Observation 4.1. □

We note that for any $\alpha < 1$, Lutz [9] has shown that the class $\text{SIZE}(\frac{2^n}{n}(1 + \frac{\alpha \log n}{n}))$ has measure 0 in ESPACE and in **C**. By the above results, this class is thus a natural example of a set that, in both ESPACE and **C**, has dimension 1 but measure 0.

**Acknowledgment.** I am very grateful to Elvira Mayordomo and John Hitchcock for useful discussions.